# Structural, optical and mechanical properties of Cr doped β-Ga$_2$O$_3$ single crystals

P. Vijayakumar [a,†], K. Ganesan [a,b,†], R.M. Sarguna [a], Edward Prabu Amaladass [a,b], M. Suganya [c], R. Ramaseshan [a,b], Sujoy Sen [a], S. Ganesamoorthy [a,b,1], P. Ramasamy [d]

[a] *Indira Gandhi Centre for Atomic Research, Kalpakkam- 603 102, India*

[b] *Homi Bhabha National Institute, Training School Complex, Anushakti Nagar, Mumbai, India*

[c] *School of Science & Humanities, Sathyabama Institute of Science and Technology, Chennai, India*

[d] *Sri Sivasubramaniya Nadar College of Engineering, Kalavakkam – 603110, Chennai, India*

**Abstract**

Undoped and Cr doped β-Ga$_2$O$_3$ (100) single crystals are grown by optical floating zone method. The full width at half maximum of rocking curve is found to be 106 arcsec for undoped Ga$_2$O$_3$ crystals whereas the 100 and 200 ppm of Cr doped Ga$_2$O$_3$ crystals display multiple rocking curves with large peak widths indicating the presence of structural defects. Raman measurements reveal broadening in the vibrational mode of ~ 350 cm$^{-1}$ with a shoulder peak indicating the Cr$^{3+}$ dopants preferentially substitute for Ga$^{3+}$ at the octahedral sites. Further, the Cr doped Ga$_2$O$_3$ crystals display strong optical absorption bands about 420 and 597 nm in the UV-Vis spectroscopy. Moreover, the observation of sharp characteristic photoluminescence emission lines at 690 and 697 nm also confirms the Cr substitution in the doped crystals. The indentation hardness increases nearly linear from 13.0 ± 0.6 to 17.9 ± 0.4 GPa whilst the indentation modulus decreases from 224.9 ± 21.4 to 202.4 ± 11.9 GPa upon Cr doping of 200 ppm in β-Ga$_2$O$_3$. The structural defects caused by the Cr doping interrupt the movement of indentation induced dislocations that results in the increase of hardness of the Cr doped β-Ga$_2$O$_3$ (100) single crystals.

---

[†] These authors contributed equally
[1] Corresponding author. Email : sgm@igcar.gov.in (S. Ganesamoorthy)

1. **Introduction**

Over the past few decades, the β-Ga$_2$O$_3$ has attracted much attention of researchers due to its ultrawide bandgap (4.9 eV), opto-electronic characteristics, chemical stability and radiation hardness. Owing to these unique physical and chemical properties, β-Ga$_2$O$_3$ is considered to be a promising candidate for wide range of applications which include high power electronics, solar blind U*V* detectors, flat panel displays and ionizing radiation detectors [1–6]. Since of β-Ga$_2$O$_3$ has high optical transparency, it is also considered as a potential host material for doping impurities with optical active centres towards optoelectronic, sensor and photonic applications [7–13]. In addition, β-Ga$_2$O$_3$ is also a fast and bright scintillator and it emits light in UV, blue and green energies, depending on the nature of defects and dopants. Even the undoped β-Ga$_2$O$_3$ crystals display UV and green luminescence due to the self-trapped excitons or lattice defects [14]. Thus, the defects play a crucial role in optical emission characteristics in undoped or doped β-Ga$_2$O$_3$. A significant amount of research on doping rare earth / transition metal in β-Ga$_2$O$_3$ exist in literature for its optical properties [4,7–14]. The trivalent chromium (Cr$^{3+}$) doping in β-Ga$_2$O$_3$ as an optical activator ion is one of the popular dopants in β-Ga$_2$O$_3$ and it emits characteristic red and near infrared emission bands with high quantum efficiency [15,16]. Moreover, the β-Ga$_2$O$_3$:Cr$^{3+}$ crystals have considerable interest in microwave and optical solid state laser applications and thin-films as electroluminescent displays [16,17].

The β-Ga$_2$O$_3$ single crystals are grown by various melt growth methods like Verneuil method, Czochralski [4], optical floating zone (OFZ) method [2,18,19], edge defined film fed growth [20], sublimation [21] and Bridgman method [22]. In Czochralski method, the main challenge is the usage of oxygen which cause oxidation of the crucible. Also, the β-Ga$_2$O$_3$ decomposes into Ga$_2$O at higher temperature under oxygen deficient atmospheres. The crucible-less OFZ technique is one of the preferred methods to grow β-Ga$_2$O$_3$ single crystals. Cr doped β-Ga$_2$O$_3$ single crystals are routinely grown by OFZ method [3,8,23]. The crystals grown under reducing atmosphere leads to oxygen deficiencies which results in n-type conductivity. Further, the incorporation of dopants into the crystal lattice introduces additional structural defects which depends on the growth parameters such as growth rate, melt temperature, rotation rate and growth environment. Intensive studies on the role of defects and their significance on the optical emission behaviour of β-Ga$_2$O$_3$ are reported [3,4,24,7–13,23] However, only a limited studies exist in literature on the mechanical properties of β-Ga$_2$O$_3$ crystals and its dependence on the defects.

The hardness and elastic modulus are fundamental properties of materials and they are directly related to crystal structure, nature of chemical bonding and an ability of the bonds to withstand deformation. Since β-$Ga_2O_3$ belongs to monoclinic crystal structure with the space group C2/m, it displays a strong anisotropy in mechanical properties. There are a few reports available on the anisotropic mechanical properties of β-$Ga_2O_3$ crystal [21,25–30]. For example, Hou et al [29] had reported the variation in the deformation mechanism in (100) and (001) planes using nanoindentation. Nikolaev et. al. [21] had reported the indentation modulus of 232 GPa for β-$Ga_2O_3$ (100) crystal. Gao et. al [30] studied the mechanical properties of (-201) plane β-$Ga_2O_3$ crystal surface with different stages of lapping and polishing steps and reported hardness value is 14.5 GPa. They concluded that the deformed layer is much harder than the perfect β-$Ga_2O_3$ crystal. The knowledge on the mechanical properties of semiconductors is crucial in the device fabrication process. Although a considerable amount of nanoindentation studies were reported on β-$Ga_2O_3$, a complete understanding on the deformation mechanism is still lacking. Overall, the reported studies on Cr doped β-$Ga_2O_3$ crystals highlight the importance of the Cr dopants on the optical emission characteristics of the host material. In addition, mechanical properties of mostly undoped β-$Ga_2O_3$ crystals and thin films are also reported independently. However, the combined studies on the role of Cr or other dopants on the mechanical properties of β-$Ga_2O_3$ are not reported. The novelty of the present work lies in the incorporation of Cr dopants in the β-$Ga_2O_3$ matrix and ensure its structural integrity using different analytical tools and also, to study how the Cr induced point defects influence on the nanomechanical properties of the grown β-$Ga_2O_3$ crystals.

In this manuscript, we investigate the growth of pure and 100 and 200 ppm $Cr^{3+}$ doped β-$Ga_2O_3$ single crystals by OFZ technique. The role of defects in the grown crystals is studied through structural and optical behavior using rocking curve analysis, Raman spectroscopy, UV-Vis optical absorption and photoluminescence (PL) emission and excitation spectroscopy. Further, the nanomechanical properties of these β-$Ga_2O_3$ crystals are studied though nanoindentation and their nanomechanical properties are corroborated with structural defects in the lattice.

2. **Experimental methods**

The pure and Cr doped (100 & 200 ppm) β-$Ga_2O_3$ single crystals were grown by OFZ using the four-mirror halogen lamp based FZ-T-4000-H-HR-I-VPO-PC OFZ crystal growth system (Crystal System Corp., Japan) and the details can be found elsewhere [31]. The (100) oriented seeds were used to grow large single crystals. The feed rods were prepared using stoichiometric compositions of high purity β-$Ga_2O_3$ (99.999%) and $Cr_2O_3$ (99.99%). The feed and seed rods were rotated in counter- clockwise direction at 30 rpm and the growth rate was varied from 3 to 7 mm/h. The diameter and length of the crystals are in the range of 6 - 8 and 30 - 40 mm respectively, depending on the growth conditions and the grown crystals are shown in Figs. 1(a-c). After the OFZ growth, dicing and polishing of the grown crystals were performed along the growth direction.

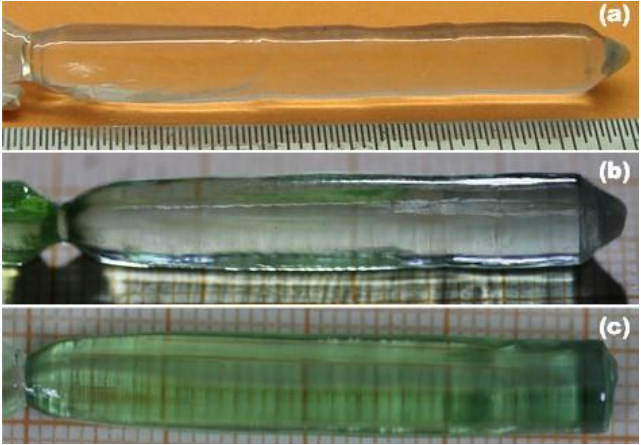

Fig.1. The as-grown β-$Ga_2O_3$ single crystals, (a) pure, (b) 100 ppm and (c) 200 ppm of Cr doping in the lattice.

The crystallinity of the grown β-$Ga_2O_3$ single crystals was examined by high resolution X-ray diffraction (HRXRD) using PANalytical multifunctional X-ray diffractometer. The well-collimated CuK$α_1$ beam was obtained from four monochromator of Ge (200) crystals set in a dispersive (+,-,-,+) configuration, which is used as an X-ray source for HRXRD measurement. Rocking curve analysis was performed through ω-scan measurement to study the low-angle grain boundaries and agglomerated point defects in the grown crystals on (100) planes. Raman spectra are recorded using micro-Raman spectrometer (Renishaw inVia) using excitation wavelength of 532 nm with the laser power of < 1 mW. The optical transmission and absorption spectra were recorded using Perkin Elmer-Lambda 35 instrument in the wavelength range of 200 - 1100 nm. The PL emission and excitation spectra are measured using FLS980, Edinburgh instruments.

Nanoindentation measurements were performed on the pure and Cr doped β-$Ga_2O_3$ wafers using a sharp three-sided pyramidal indenter (Berkovich type). The hardness of the β-Ga$_2$O$_3$ crystal wafers is extracted from load-depth curves using Oliver-Pharr method (H$_{OP}$). The single-cycle load–displacement curves (P–h curves) were measured at the load of 1 mN. As nanoindentation measurements are usually performed in the nano scales, even small defects in the indenter tip may lead to large differences in measurement results. Hence, the 3 × 3 matrix indents were made at different locations of its crystal wafers and an average hardness value was obtained. The elastic indentation modulus (Young's modulus) were also obtained on each cycle of loading and unloading curve using Oliver and Pharr method.

## 3. Results and Discussion

### 3.1. High resolution X-ray diffraction

Fig. 2a shows the Laue diffraction pattern of pure β-Ga$_2$O$_3$ crystal. The sharp and bright diffraction spots confirms the single crystalline nature. Figs. 2(b-d) display the rocking curves recorded for pure and doped β-Ga$_2$O$_3$ crystal wafers on (100) diffracting planes. As shown in the Fig. 2(b-d), the rocking curve in pure β-Ga$_2$O$_3$ crystal wafer is having a single diffraction peak with full width at half maximum (FWHM) of 104 arc sec. However, the Cr doped crystals have multiple diffraction peaks around the (100) plane that indicate the presence of structural defects in the form of low angle grain boundaries or micro-twining [32]. The rocking curves were deconvoluted by Lorentz fit to investigate the additional peaks. In the 100 ppm Cr doped β-Ga$_2$O$_3$ crystal, a total of four diffraction peaks are found with FWHM of ~ 404, 457, 259 and 1073 arc. sec. in the ω-scan range of -1.5 to 1.5° On the other hand, a total of five diffraction peaks with FWHM of 510, 498, 514, 845 and 487 arc. sec are observed for the 200 ppm Cr doped β-Ga$_2$O$_3$ crystal. The number of additional rocking curves and their FHHM increase with Cr doping concentration in β-Ga$_2$O$_3$ lattice. This observation indicates the increase in micro-twining / other structural disorder with Cr doping in the β-Ga$_2$O$_3$ matrix as compared to that of undoped crystal [32]. The difference in the ionic radius of Cr dopant induces strain in the lattice that results in low angle grain boundaries or micro-twins. Further, the large temperature gradient during the OFZ growth process may also be responsible for the formation of these low angle grain boundaries.

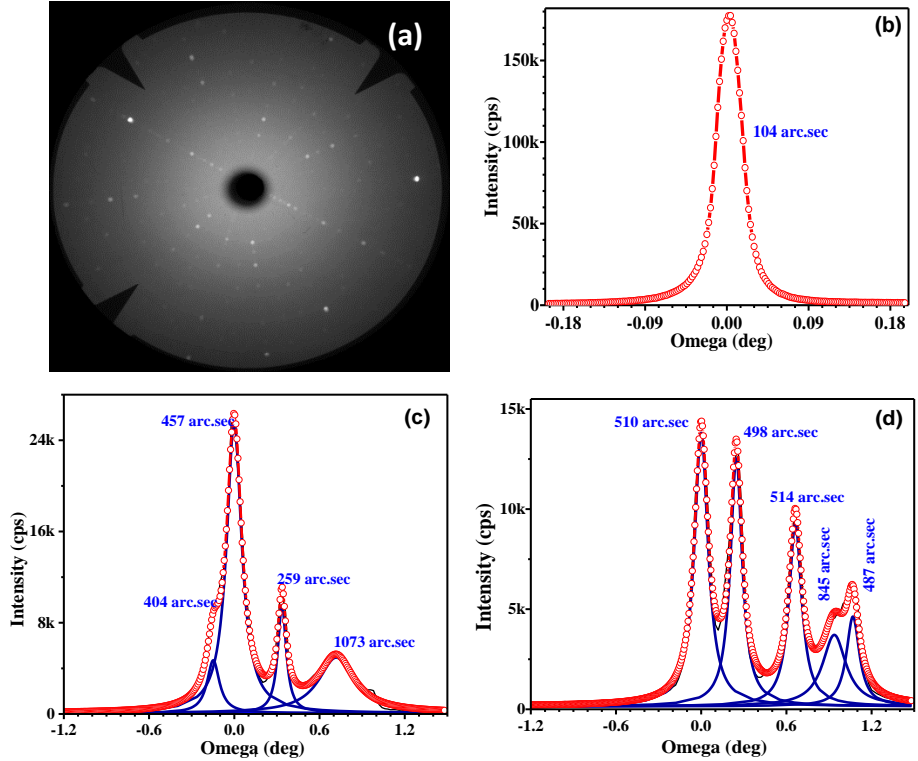

Fig. 2. (a) Laue diffraction pattern of (100) pure β-Ga$_2$O$_3$ single crystal. The X-ray rocking curve on (b) pure, (c) 100 ppm and (d) 200 ppm of Cr doped β-Ga$_2$O$_3$ single crystal

## 3.2. Raman Analysis

Fig. 3 shows the Raman spectra of pure and Cr doped β-Ga$_2$O$_3$ single crystals. Here, the spectra are stacked vertical and the intensity is normalized with A$_g$(3) mode for comparison. The β-Ga$_2$O$_3$ unit cell is composed of distorted [Ga$_I$O$_4$] tetrahedra and [Ga$_{II}$O$_6$] octahedral, with three types of oxygen (O) ions and two types of gallium (Ga) ions. The tetrahedral and octahedral chains are aligned along the b-axis. According to the factor group analysis, there are fifteen (10A$_g$ and 5B$_g$) Raman active optical zone center phonon modes in β-Ga$_2$O$_3$ [33]. Among them, fourteen Raman active modes are observed and the modes are labelled in Fig. 3. The Raman modes at low frequency region (< 200 cm$^{-1}$) arise from the vibration of lattice as a whole associated with tetrahedral chains; the mid frequency modes (300-500 cm$^{-1}$) are from the deformation of the tetrahedral and octahedral chains; and the high frequency modes (> 500 cm$^{-1}$) originate from the stretching and bending of Ga$_I$O$_4$ tetrahedrons [34]. The observed peak position of Raman modes of pure, and Cr doped β-Ga$_2$O$_3$ do not vary significantly as can be evidenced from Fig. 3. However, the FWHM of some of the Raman modes increase slightly for 200 ppm of Cr doped in β-$Ga_2O_3$. In addition, the relative intensity of Raman modes ($A_1$, $A_2$, $B_1$, $B_2$, $A_5$, $A_6$ and $A_{10}$) also decrease significantly as can be seen in Fig. 3. Especially the intensity of $A_{10}$ mode decreases drastically upon Cr doping. The notable change in the Raman spectra of pure and doped $Ga_2O_3$ crystals is that the broadening of the A5 mode ~ 346.50 cm$^{-1}$ and the appearance of an additional peak at higher wavenumber (370.3 cm$^{-1}$). This result can be explained by considering that $Cr^{3+}$ atoms preferentially substitute for $Ga^{3+}$ at the octahedral sites which deformed the tetrahedra structure locally [16]. Hence, the substitution of Cr as a dopant results in the appearance of higher energy shoulder for the 346.50 cm$^{-1}$ peak. Also, the insignificant change in FWHM of Raman bands with Cr doping in $Ga_2O_3$ indicating the overall structural integrity of the grown crystals.

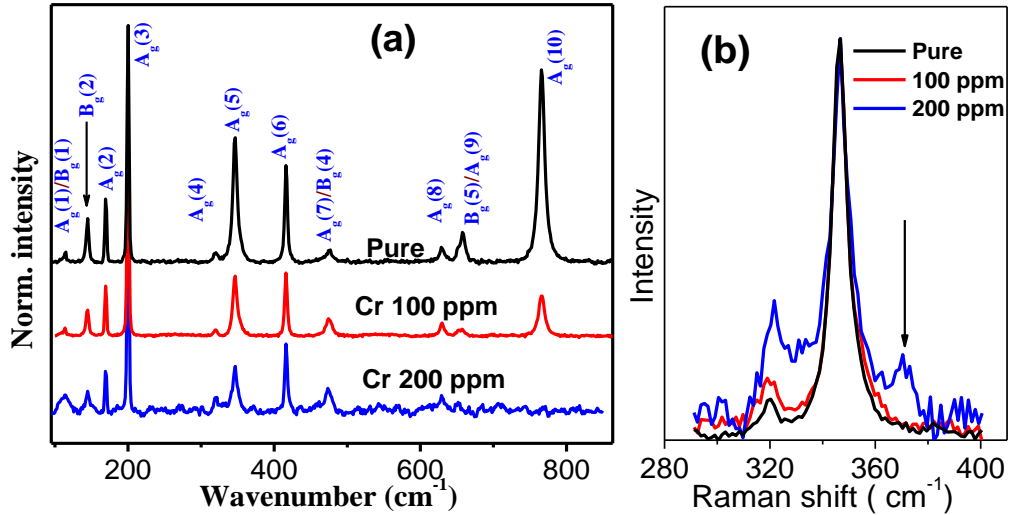

Fig.3. (a) Raman spectra for pure and Cr doped $Ga_2O_3$ crystals and (b-e) magnified and normalized Raman spectra of selected phonon mode

### 3.3. UV-Vis absorption spectroscopy

Fig. 4a depicts the UV-vis-NIR transmission spectra for pure and Cr doped $Ga_2O_3$ crystals. The spectra are recorded under identical condition with wafer thickness of ~ 0.6 mm. A sharp transmission cut off wavelength are observed at 262, 263 and 265 nm for pure, 100 and 200 ppm of Cr doped single crystals, respectively. The undoped $Ga_2O_3$ single crystals display transmittance of ~ 80 % in the visible-NIR wavelength range while the transmittance of doped crystals have decreased slightly in visible-NIR region due to Cr doping. Further, the Cr doped $Ga_2O_3$ crystals have two absorption bands centred at 420 and 597 nm which correspond to $^4A_2 \rightarrow {}^4T_1$ and $^4A_2 \rightarrow {}^4T_2$ transitions, respectively. The optical band gap of the grown crystals is calculated using the Tauc plot,

$$\alpha = 2.303 \log(100/T)/d \qquad \text{...................................... (1)}$$

where α is the absorption coefficient, T is transmission in percentage and d is thickness of the wafer. Fig. 4b shows a plot of the relative absorption coefficient $(\alpha h\nu)^2$ versus the incident photon energy, hν. The estimated optical band gap energy values are 4.64, 4.63, and 4.60 eV for pure, 100 and 200 ppm Cr doped β-$Ga_2O_3$ single crystals, respectively. The observed bandgap values of the grown crystals are comparable to the reported data in literature [8,35]. The incorporation of Cr ions at the Ga site of β-$Ga_2O_3$ contributes to new electron states within the forbidden energy gap that leads to a slight decrease in the bandgap energy.

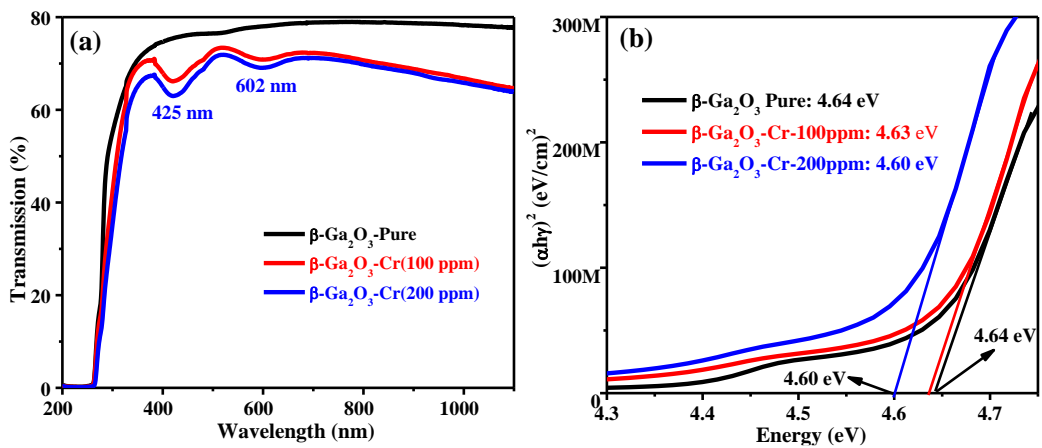

Fig.4. (a) UV-vis optical transmission spectra, and (b) Tauc plot for pure and Cr doped $Ga_2O_3$ wafers

### 3.4. Photoluminescence spectroscopy

Fig. 5a depicts PL emission spectra of pure and Cr doped $Ga_2O_3$ crystals that are excited at 260 nm. The pure β-$Ga_2O_3$ exhibits a broad PL emission band in the UV region of about 385 nm with a shoulder peak about 430 nm. Here, the 385 and 430 nm emission bands are associated with oxygen vacancies in the lattice and they are assigned to recombination of a self-trapped excitons and a donor–acceptor pair, respectively [36–38]. Further, the intensity of the 385 nm emission line decreases drastically with addition of $Cr^{3+}$ ions in the lattice. Moreover, there is a blue shift for the 385 nm defect band which appears at ~ 375 nm for 100 ppm Cr doped $Ga_2O_3$. Moreover, this defect band is almost disappeared for 200 ppm $Cr^{3+}$ doped β-$Ga_2O_3$. Here, the suppression of this UV emission in Cr doped β-$Ga_2O_3$ crystals indicates the non-radiative behaviour of $Cr^{3+}$ ions for photo-excited carriers.

The PL emission spectra also exhibit sharp characteristic emission lines at 690 and 697 nm in Cr doped in β-Ga2O3. The narrow red emission is ascribed to the electronic transition of $Cr^{3+}$ having the $d^3$ configuration from the excited $^4T_2$ level to the ground state of $^4A_2$ level. Moreover, the intensity of these characteristic emission lines and broad background emission band ~ 720 nm increases with Cr concentration in the crystal. In addition, a broad green emission band is also observed at ~ 525 and 550 nm for 100 and 200 ppm of $Cr^{3+}$ doped

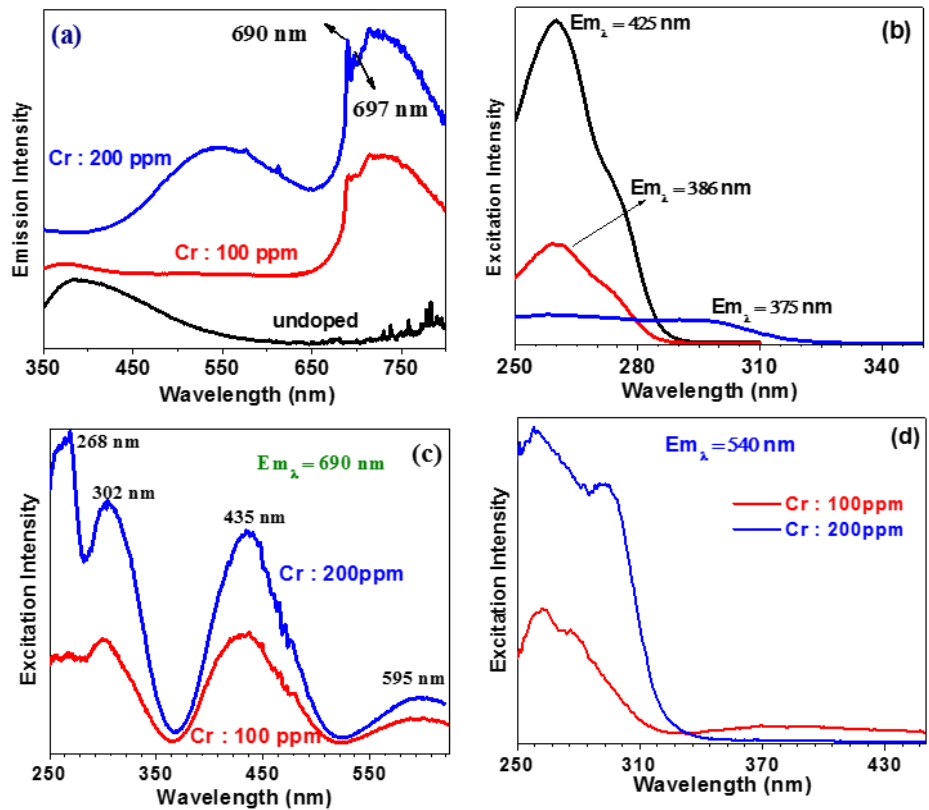

Fig.5. (a) Photoluminescence emission spectra of pure and Cr doped β-Ga2O3 single crystals with excitation wavelength of 255 nm. (b-d) Photoluminescence excitation intensity of pure and Cr doped β-Ga2O3 for different emission wavelengths. β-Ga2O3 crystals, respectively. The intensity of green emission is also enhanced upon increase in the $Cr^{3+}$ concentration. The green emission of undoped β-Ga2O3 is associated with the energy transfer between excitons to oxygen related defects [14].

Figs. 5b-5d show the PL excitation (PLE) spectra of the $Cr^{3+}$ doped *β*-Ga2O3 crystals for different emission peaks. The peak wavelength of the excitation band around ~ 262 nm coincides with the band gap of respective pure and Cr doped *β*-Ga2O3 crystals. However, the excitation spectra differ in structure between the green and the red emission of $Cr^{3+}$ doped β-

Ga$_2$O$_3$ as can be seen from Figs. 5c and 5d. The excitation peaks ~ 435 and 595 nm are due to the charge transfer between the states $^4A_2$ to $^4T_1$ and $^4A_2$ to $^4T_2$ level, respectively [39]. These observed optical behaviour demonstrate the high structural quality of the β-Ga$_2$O$_3$ crystal, consistent with Raman analysis, and also the Cr doped β-Ga$_2$O$_3$ is a good host material to produce characteristic luminescence emission from transition metal ions under UV excitation.

### 3.5 Mechanical properties

Fig. 6a shows the typical plot of loading – unloading curves ie. load vs displacement curves for pure and Cr doped β-Ga$_2$O$_3$ crystals. The mean of maximum displacement decreases from 57.3 ± 1.7 nm to 51.4 ± 1.2 nm upon Cr doping in β-Ga$_2$O$_3$ crystals at the load of 1 mN. The decrease in penetration depth indicates that the doped crystals have better rigidity for applied load as compared to that of pure crystal. Further, a careful observation on the load – displacement curve indicates several pop-in events on these crystals. One such pop-in event is magnified and shown as inset in Fig. 6a. The pop-in event is a signature of sudden displacement jump which is typically be associated with the evolution of dislocation and cracks in the lattice [40]. Fig. 6b shows the variation of indentation hardness and indentation modulus of β-Ga$_2$O$_3$ (100) crystal as a function of Cr doping concentration. The indentation hardness of the crystals is found to be ~ 13.0 ± 0.6, 15.1 ± 1.0, and 18.0 ± 0.4 GPa for 0, 100 and 200 ppm of Cr in β-Ga$_2$O$_3$, respectively. The increase in hardness of β-Ga$_2$O$_3$ with Cr doping is attributed to the increase in structural defects that are associated with Cr doping. Such defects impede the movement of slip planes / dislocation during the indentation and hence, the increase in hardness. Moreover, the increase in structural disorder also corroborates with XRD rocking curve analysis.

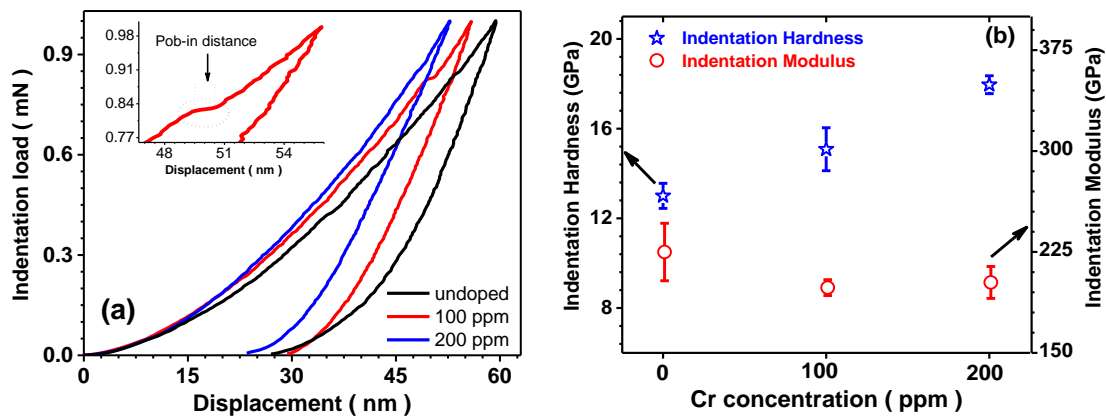

Fig. 6. (a) Typical load versus displacement curves and, (b) indentation hardness and indentation modulus of pure, 100 and 200 ppm of Cr doped β-Ga$_2$O$_3$ crystals at nanoindentation load of 1 mN. The inset shows the magnified part of the load versus displacement curve for 100 ppm Cr doped β-Ga$_2$O$_3$ crystal.

The highest indentation hardness of 18 GPa is observed for 200 ppm Cr doped β-$Ga_2O_3$ crystal which had larger defects than that of 100 ppm of Cr doped β-$Ga_2O_3$ crystal (15 GPa). The indentation pure $Ga_2O_3$ wafers hardness is still lower as compared to Cr doped β-$Ga_2O_3$ crystal. On the other hand, the indentation modulus of the pure crystal is ~ 225.0 ± 21.4 GPa and it decreases downto ~ 198.5 ± 5.9 GPa for 100 ppm and ~ 202.4 ± 11.9 GPa for 200 ppm Cr doped β-$Ga_2O_3$ crystal. The observed elastic modulus of ~ 225 GPa is very close to the reported value of ~ 232.0 Gpa for pure (100) β-$Ga_2O_3$ crystal [21]. Here, the reduction in elastic modulus of Cr doped β-$Ga_2O_3$ crystals is attributed to the decrease in packing fraction which arises due to the structural disorder caused by Cr doping in the lattice.

## 4. Conclusions

The pure and Cr doped β-$Ga_2O_3$ single crystals were grown by four mirror optical floating zone technique. The growth parameters were optimized and the crystals were grown along (100) orientation. The Cr dopants in β-$Ga_2O_3$ introduced certain structural disorder in the lattice as evidenced by rocking curve analysis. The high optical transparency of ~ 70 – 80 % in visible-NIR region confirms the high structural quality of grown β-$Ga_2O_3$ single crystals. The $Cr^{3+}$ doping causes strong absorption bands about 420 and 597 nm in the UV-Vis transmission spectra. The $Cr^{3+}$ doping also induces sharp characteristic emission lines at 690 and 697 nm as probed by photoluminescence spectroscopy. Further, the optical properties of the β-$Ga_2O_3$ single crystals are significantly enhanced with Cr doping while the structural properties are still intact. Moreover, the Cr dopants increase the hardness from 13 to 18 GPa of β-$Ga_2O_3$ crystals while retaining its good structural and optical properties. This study concludes that the optical floating zone technique is capable to grow high quality β-$Ga_2O_3$ single crystals.

## 5. Acknowledgement

**Author contribution**: P. Vijayakumar -  Investigation, Methodology, Formal analysis, Data curation, Writing- Original draft; K. Ganesan – Investigation, Formal analysis, Writing- Review & editing; R.M. Sarguna, Edward Prabu Amaladass, M. Suganya,  Sujoy Sen – Investigation; R. Ramaseshan - Investigation, Writing- Review & editing; S. Ganesamoorthy - Conceptualization, Supervision, writing – Review & editing, P. Ramasamy – Resources, Writing – Review & editing

**Data availability** : Data of this study are available from the corresponding author on reasonable request.

**Declarations**

Conflict of interest : The authors declare that they have no conflict of interest.